\documentstyle[11pt,paspconf]{article}
\input psfig.tex

\def\lsim{\lower.5ex\hbox{$\; \buildrel < \over \sim \;$}}
\def\gsim{\lower.5ex\hbox{$\; \buildrel > \over \sim \;$}} 

\begin{document}

\title{ACCRETION PROCESSES AROUND BLACK HOLES AND NEUTRON STARS: ADVECTIVE DISK PARADIGM}
%\altaffilmark{1}}

\author{S. K. Chakrabarti\altaffilmark{1}}
\affil{S.N. Bose National Center For Basic Sciences, JD-Block, Salt Lake, Calcutta-700091}

\altaffiltext{1}{email: chakraba@bose.ernet.in}

\begin{abstract}
We review models which include advective accretion disks onto compact 
objects and discuss the influence of the centrifugal pressure supported high 
density region close to the compact objects on the emitted spectra. 
We show that the stationary and non-stationary spectral properties 
(such as low and high states, quasi-periodic 
oscillations, quiescent and rising phases of X-ray novae, etc.) of 
black hole candidates could be satisfactorily explained by the 
advective disk models. 
\end{abstract}

\section{Introduction}

Physics of accretion has undergone major revisions in every twenty years or 
so. In the 1950s, spherical accretion onto compact stars was studied rather 
extensively. This so-called Bondi flow (Bondi 1952) is radiatively very 
inefficient as matter has a very high radial velocity and therefore, for a
given accretion rate, it has a very low density. When quasars and active 
galactic nuclei (AGN) were discovered, it was impossible to explain their 
luminosity by using the properties of the Bondi flow, and several workers 
engaged in finding ways to increase the efficiency of the flow by addition 
of magnetic field and plasma processes (see, Chakrabarti 1996a [C96a] for a 
recent review). Partially to resolve the efficiency problem, and partially 
otherwise, in the 1970s, another extreme form of accretion, namely, 
the Keplerian disk models, became popular thanks to the pioneering works 
of Shakura \& Sunyaev (1973, [SS73]) and Novikov \& Thorne (1973). 
Here the disk is always assumed to have the Keplerian 
angular momentum distribution and the flow pressure 
and advection effects were assumed to be negligible. The energy 
generated locally by viscous dissipation is assumed to be totally radiated 
away. Although the inner edge of the disk is assumed to coincide 
with the innermost stable circular orbit and no attempt was made 
to satisfy inner boundary condition on the horizon, this disk model
nevertheless, was {\it very} successful in explaining the `big blue
bump' observed in the active galaxies as well as the soft X-ray
bump observed in low mass binary systems. However, the galactic black hole
candidates were soon found to appear in roughly two distinct
states, one is the soft state and the other is the hard state (see, Ebisawa,
Titarchuk \& Chakrabarti, 1996 [ETC96] for a list of known candidates and 
their observed states). In soft states, more power is emitted in soft X-rays
and the multicolor-black body bump comes along with a weak power
law component of slope $\alpha \sim 1.5$ ($F(\nu) \sim \nu^{-\alpha}$).
In hard states, on the contrary, more power is emitted in hard X-rays;
soft bump is faint or nonexistent, and the power law component has a slope 
of $\alpha \sim 0.5$. Such observations clearly challenged the simplistic
Keplerian disk model and a further revision of the disk model was necessary.

\section {Accretion Flows of 1990s: The Advective Disks}

Not surprisingly, theoreticians sensed the inadequacies of the standard
Keplerian disk models in early 1980s. Paczy\'nski \& Bisnovatyi-Kogan
(1980) wrote down the so-called advective disk equations which 
contained the advection and pressure terms. For optically thick
disks, solutions were being tried out under various assumptions. 
In slim disk model (Abramowicz et al., 1988) it was shown that 
(locally) the disk no longer has the viscous and thermal instability
when advection was added, but the global solution was found to be incorrect
as the angular momentum was seen to deviate from a Keplerian distribution
far away from the black hole. First completely global solutions of the viscous,
advective disks were presented in Chakrabarti (1990a, [C90a]) where it was 
assumed that the heating and cooling were so adjusted that the flow would be 
isothermal. These assumptions were removed afterwards (Chakrabarti 1996b,
[C96b]) and the conclusions about the topology of the solutions remained the 
same. Both optically thin and optically thick single temperature
solutions were obtained. Generally speaking, the fundamental 
points in an advective disk is the following:

Matter entering a black hole must possess radial velocity comparable to the
velocity of light on the horizon and as a result the flow must be super-sonic
and sub-Keplerian (Chakrabarti 1990b [C90b]; C96ab). The advective flow 
deviates from a Keplerian disk away from the black hole depending on viscosity
and cooling/heating processes and eventually passes through a sonic point 
before entering the hole. If the flow is hot enough (or, away from equatorial 
plane), it may also pass through a standing shock and subsequently, 
through a second (inner) sonic point. On a neutron star, on the other hand,
the flow may remain subsonic throughout, or, if it ever becomes super-sonic, 
it must pass through a shock at the outer edge of the boundary layer.
This is because the radial flow must stop due to pressure of the
radiation emitted from the surface and hence the flow must be sub-sonic.
A thin boundary layer is produced within which both the rotational
velocity and radial velocity reach the star-surface values. 

%{\parfillskip=0pt 
The formation of a centrifugal pressure supported standing shock
around a black hole comes about because of the following reason: close 
to the black hole,
the infall time scale of the flow is usually very small compared to the 
time scale of viscous transport of angular momentum. As a result, angular 
momentum remains almost constant in last tens of Schwarzschild radii 
(if the viscosity is small enough) and the centrifugal force increases 
faster compared to the gravitational force. Matter piles up behind the 
centrifugal barrier and a shock forms. The shock need not be sharp, i.e.,
the jump in temperature and density may be gradual, depending on the viscosity
of the flow. The important point is that the flow slows down close to the
hole and forms a dense ionized cloud of hot gas. Farther close to the hole, 
the flow passes through the inner sonic point and enters into the hole
supersonically. When the viscosity is high, the centrifugal
barrier is absent altogether and Keplerian disk would be present till close 
to the marginally stable orbit. This approach of the Keplerian disk towards 
the hole is gradual as the extent of viscosity (and therefore accretion rate 
of the Keplerian component) is increased. This phenomena
is a part of solutions of the viscous transonic flow (C96b)
and has been confirmed by XTE observations rather adequately (Zhang et al.,
1997).

Given the above description of the solution (which will be 
quantified below), one can imagine that the most general 
nature of accretion disk that could form around a black hole 
would consist of several components as shown in Fig. 1. This picture is 
also valid in wind fed binary systems as well as in active galactic nuclei
where the stellar winds are also accreted. The inherent assumption
that is involved in such a picture is that the viscosity parameter (SS73)
is decreasing away from the equatorial plane and that the angular momentum
and the inner sonic point close to the black hole are roughly similar
at all heights. High viscosity advective flows near the equatorial plane 
continue to form a Keplerian disk all the way close to the
black hole, but the low viscosity (with alpha parameter $\sim 0.05$)
flows deviate from a Keplerian disk and has the centrifugal pressure 
supported dense region close to the black hole. Flow with even smaller 
viscosity $\alpha \lsim 0.001$) may form a gigantic extended atmosphere 
which is basically rotating as an ion torus of size $\sim 10^{3-4}x_g$, where 
$x_g=2GM/c^2$ is the Schwarzschild radius of the central black hole of mass $M$.
%\par}
\begin{figure}
\vbox{
\vskip -3.0cm
\centerline{
\psfig{figure=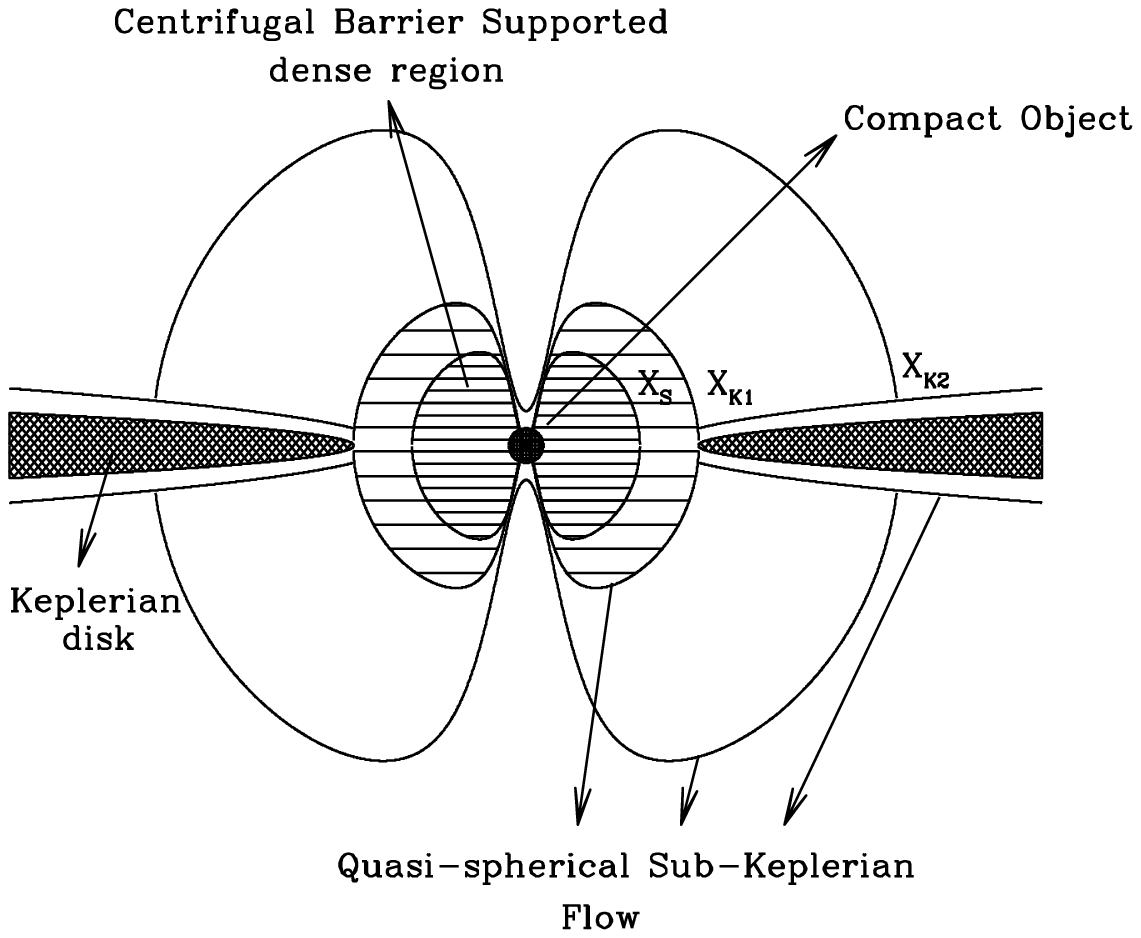,height=15.0truecm,width=15.0truecm,angle=0}}}
\vspace{-5.0cm}
\noindent {\small {\bf Fig. 1:} 
Schematic diagram of the multi-component advective disk model. 
Keplerian disk component which eventually becomes sub-Keplerian 
close to a compact object is flanked by a sub-Keplerian halo 
component which originates from the Keplerian disk farther out 
(depending on viscosity) and possibly  contributed by winds of the 
companion or nearby stars. Giant predominantly rotating disk of size 
$\sim 10^{3-4} x_g$ forms an extended atmosphere which reprocesses radiations
emitted from inner disks. }
\end{figure}

Fig. 2 shows the classification of the {\it entire} parameter space
according to the types of solutions of inviscid flows
around a black hole (Chakrabarti 1996c [C96c]; see also Chakrabarti 1989 
[C89]; C90b). 
In the central box, the parameter space (spanned by specific 
angular momentum $l$ and specific energy ${\cal E}$) is divided
into nine regions marked by $N$, $O$, $NSA$, $SA$, $SW$, $NSW$, $I$,
$O^*$, $I^*$. The horizontal line at ${\cal E}=1$ corresponds to the rest
mass of the flow. Surrounding this parameter space, 
\begin{figure}
\vbox{
\vskip -1.0cm
\centerline{
\psfig{figure=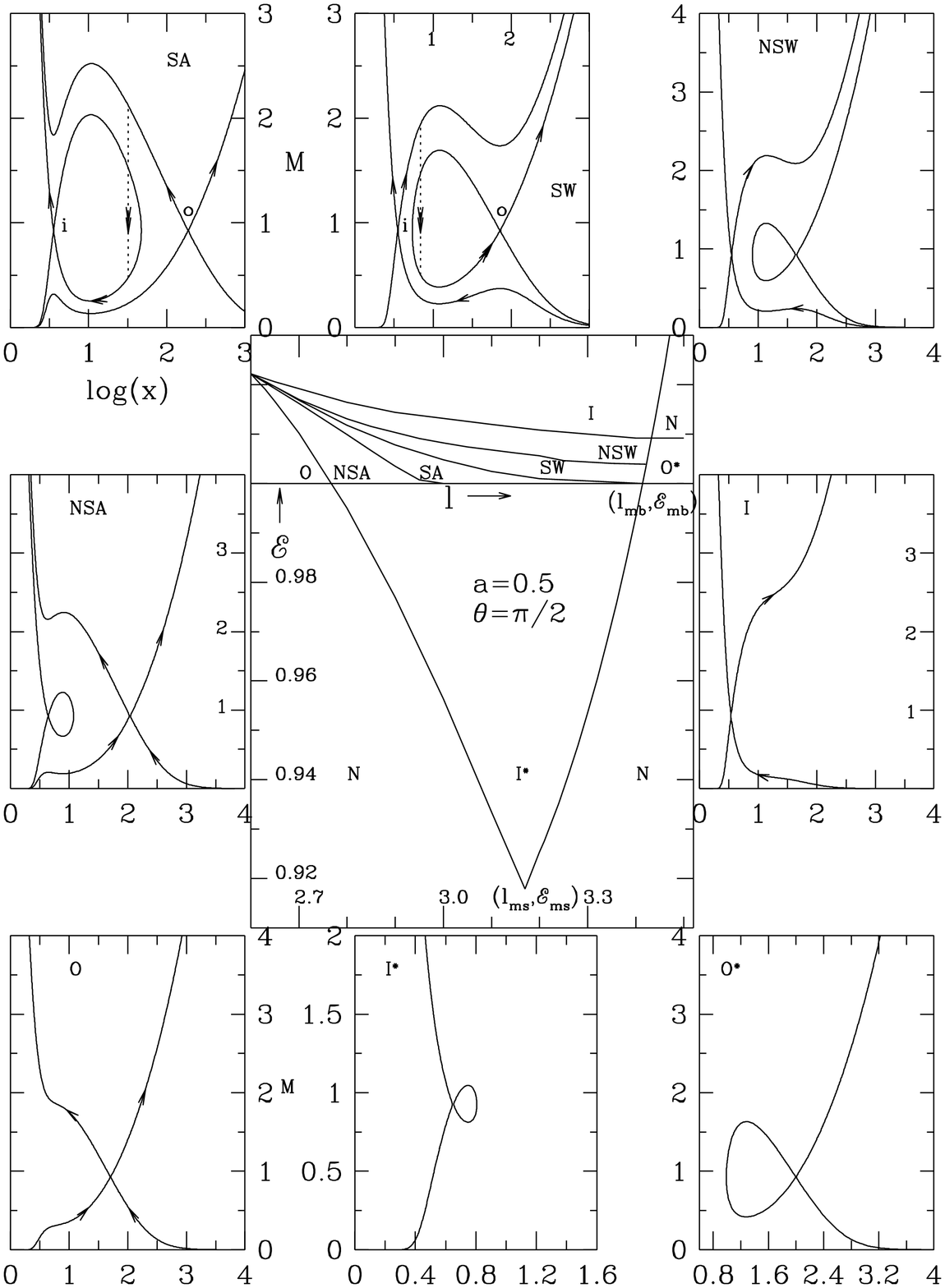,height=11.0truecm,width=11.0truecm,angle=0}}}
\vspace {0.5cm}
\noindent {\small {\bf Fig. 2:} 
Classification of the {\it entire} parameter space (central box)
in the energy-angular momentum plane in terms of topological variation 
of the Kerr black hole accretion ($a=0.5$ and polytropic index $\gamma=4/3$). 
Eight surrounding boxes show the solutions from each independent region of the
parameter space. Contours are of constant entropy accretion rate 
${\dot{\cal M}}$. Arrowed curves are the solutions which pass through
sonic points. Vertical arrowed lines correspond to shock transitions.}
\end{figure}
%\vspace{8.0cm}
%{\parfillskip=0pt 
various solutions (Mach number $M=v_x/a_s$ vs. logarithmic radial distance $x$
where $v_x$ is the radial velocity and $a_s$ is the sound speed) marked
with the same notations (except $N$) are plotted. The accretion solutions have 
inward pointing arrows and the wind solutions have outward pointing arrows.
The region $N$ has no transonic solution. $E$ and $l$ are the only two 
parameters required to describe the entire inviscid global solutions. Since 
$E$ is assumed to be constant, entire energy is advected towards the hole. 
Thus, these solutions are hot but inefficient radiators very similar to their
spherical counterpart (Bondi flow). The constancy of energy is roughly
guaranteed in optically thin solutions only. Modifications of these solutions
in viscous flows which include heating and cooling are in C96b. In the case 
of neutron star accretion, the subsonic inner boundary condition forces the 
flow to choose the sub-sonic branch and therefore the energy must be 
dissipated at the shock (at $x_{s1}$ or $x_{s3}$ in the notation of C89)
outside the neutron star surface (C89; C90b; C96b)
unless the entire flow is subsonic. This is the essential difference
between black hole and neutron star accretions: In the black hole
accretion the total luminosity could be much less compared to 
the maximum luminosity permitted by general relativistic
considerations (C96a) since the rest could be
advected through the horizon (as we see here, perfectly
stable solutions exist even with constant specific energy), 
while in neutron stars (at least when the magnetic field is absent) 
the matter must emit all radiations outside the star surface.
The solutions from the region `O' has only the outer sonic point. 
The solutions from the regions $NSA$ and $SA$ have two `X' type sonic points
with the entropy density $S_o$ at the outer sonic point {\it less} than the
entropy density $S_i$ at the inner sonic point. However, flows from $SA$
pass through a standing shock  since the Rankine-Hugoniot
conditions are satisfied. The entropy generated at the shock 
$S_i-S_o$ is advected towards the black hole to enable the flow to pass
through the inner sonic point. These advective disk 
solutions have been verified to be {\it stable}
by detailed numerical simulations (Chakrabarti \& Molteni, 1993;
Molteni, Lanzafame \& Chakrabarti, 1994; Molteni, Ryu \& Chakrabarti, 1996).
Rankine-Hugoniot conditions are not 
satisfied for flows from the region $NSA$. Numerical simulations show
(Ryu, Chakrabarti \& Molteni, 1997 [RCM97])
that flows from this region are very unstable
and exhibit periodic changes in emission properties as they
constantly try to form stationary shocks, but fail to do so. The frequency 
and amplitude of modulation (10-50\%) of emitted X-rays have properties similar
to Quasi-Periodic Oscillations (QPOs) observed in black hole candidates 
(Dotani, 1992). In galactic black holes, these frequencies are around 1Hz 
(exact number depends on shock location, i.e., $l,\ {\cal E}$ parameters) but 
for extragalactic systems the time scale could range from a few hours
to a few days depending on the central mass ($T_{QPO} \propto M_{BH}$).
Numerous cases of QPOs are reported in the literature 
(e.g., Dotani, 1992; Halpern \& Marshall, 1996; Papadakis \& Lawrence, 1995).
In presence of cooling effects, otherwise stationary shocks from $SA$ also 
oscillate with frequency and amplitude modulations comparable to those of
QPOs {\it provided} the cooling timescale is roughly comparable to the
infall timescale in the post-shock region (Molteni, Sponholz \& Chakrabarti,
1996, [MSC96]). Kilohertz oscillations on neutron stars 
are also possible when the shock at $x_{s1}$ form (typically,
at $2.5 x_g$, just outside the neutron star surface). This is because the
Comptonization time scale and the infall time scale are both
comparable to $\sim 0.001s$ (Chakrabarti \& Titarchuk, in preparation).
The solutions from the regions $SW$ and $NSW$ are very similar to those from
$SA$ and $NSA$. However, $S_o \geq S_i$ in these cases.
Shocks can form only in winds from the region $SW$. Shock conditions are not
satisfied in winds from the region $NSW$. This makes the $NSW$ flows
unstable as well. A flow from region $I$ has only the inner sonic
point and thus can form shocks (which require the presence of two 
saddle type sonic points) if the inflow is already supersonic 
due to some other physical processes (as in a wind-fed system). Each solution 
from regions $I^*$ and $O^*$ has two sonic points (one `X' and one `O')
only and neither produces complete and global solution. The region $I^*$
has an inner sonic point but the solution does not extend subsonically
to a large distance. The region $O^*$ has an outer sonic point, but the
solution does not extend supersonically
to the horizon! When a significant viscosity is added, the closed
topology of $I^*$ opens up and then the flow joins with a cool Keplerian 
disk (C90ab; C96b) which has ${\cal E} <1$. These special solutions of viscous 
transonic flows should not have shock waves. However, hot flows deviating from
a Keplerian disk or sub-Keplerian companion winds, or flows away from
an equatorial plane (C96d) 
or, cool flows subsequently preheated by magnetic flares or irradiation
can have ${\cal E}>1$ and therefore standing shock waves. 
Note that in order to have standing shocks, one does not
require large angular momentum. In majority of the cases, the flow
needs to have $l<<l_{ms}$, the marginally stable value (Fig. 2). Although 
an adiabatic flow with polytropic index $\gamma<1.5$ does not have shocks, 
the stationary observational properties, which depend only on the 
enhanced emission from the region behind of centrifugal barrier, are not 
affected.
%\par}
%\newpage
\begin{figure}
\centerline{
\psfig{figure=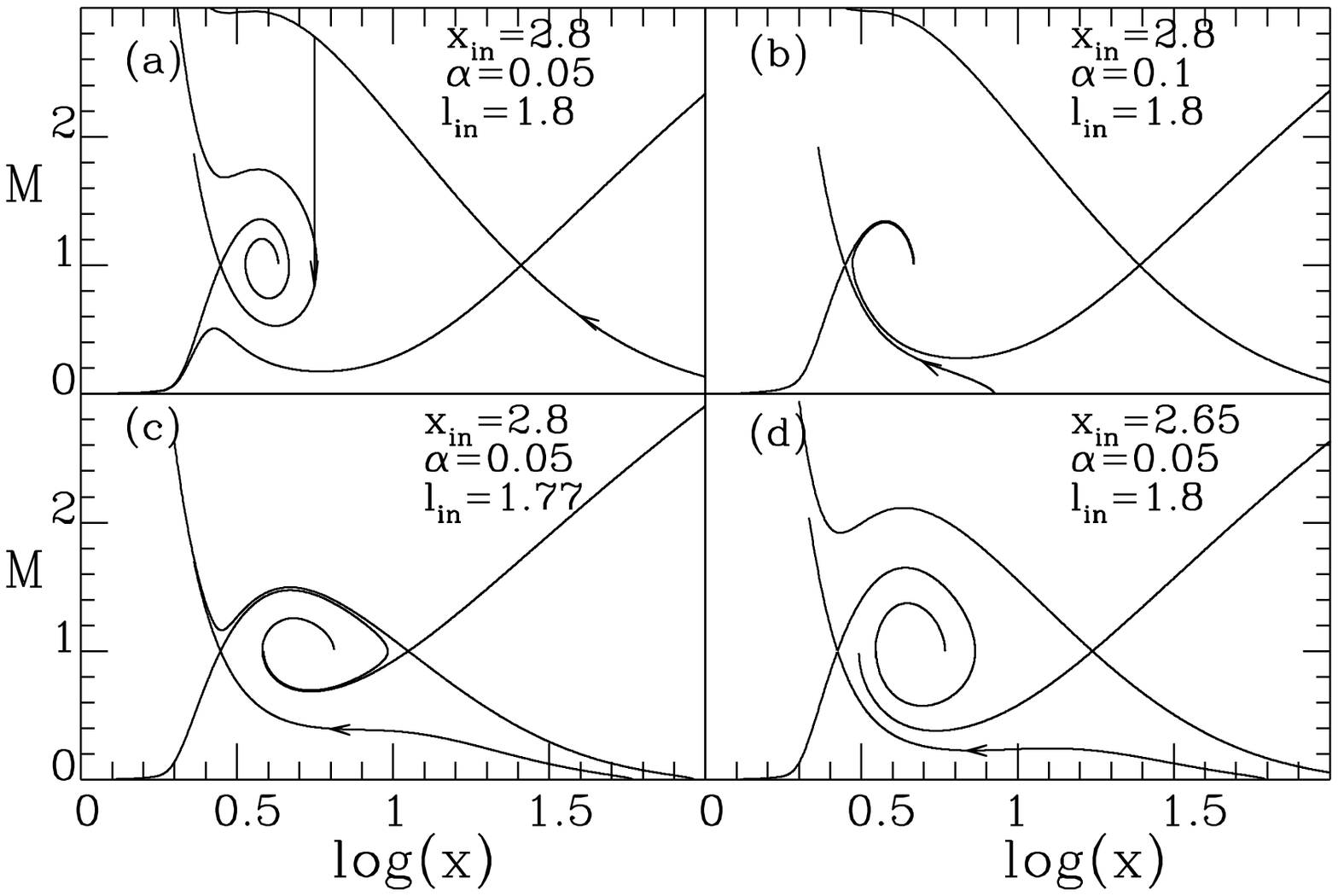,height=13.0truecm,width=13.0truecm,angle=0}}
\vspace{-7.0cm}
\noindent {\small {\bf Fig. 3:} 
Change in solution topology of a viscous flow as all three parameters are 
varied (C90a). In (a), after deviating from a hot Keplerian 
disk the flow can pass through  the outer sonic point, 
shock and the inner sonic point (shown in arrows), while
in (b-d) the stable flow can pass through the inner sonic point after it
deviates from a Keplerian disk. Density and velocity distributions in the 
flow close to the black hole remain roughly the same in all these cases. 
Distance $x$ is measured in units of $x_g$.}
\end{figure}
%\vspace{7.0cm}

When viscosity is added, the closed topologies shown in  
Fig. 2 open up as the `O' type sonic points become spiral or nodal type.
Singularly important in this context (perhaps the most important
understanding in the accretion disk physics since Bondi flow!) 
is the non-trivial change in topology 
when each of the three free parameters are changed (C90ab). 
In the context of viscous isothermal flows (These discussions are valid
for a general flow as well, see, C96ab.)
these free parameters can be chosen to be inner sonic point $x_{in}$ 
(this replaces the specific energy parameter), the angular momentum 
on the horizon $l_{in}$ and the viscosity parameter $\alpha$.
Temperature of the disk is computed self-consistently from these parameters.
Fig. 3 shows that transition to topology in (b-d) from topology (a) 
can take place either by increase in viscosity or decrease 
$x_{in}$ or $l_{in}$. In (a), shocks are still possible, while in (b-d),
shocks do not form as the flow enters into the hole through the inner sonic 
point straight away from a Keplerian disk, but the density variation and 
emission properties remain similar to that of a shocked flow. In (b),
Keplerian disk is extended close to the horizon, while in (a)
the deviation takes place farther outside of the outer sonic point. Thus, 
for instance, if there is a vertical variation of viscosity in a Keplerian 
disk very far 
away from a black hole, it is possible that different layers would deviate 
from a Keplerian disk at different radial distance, and a sub-Keplerian flow
(with or without a shock) would surround a Keplerian disk as the flow
approaches the compact object as shown in Fig. 1.
This two-component advective flow (TCAF) Model (Chakrabarti \& Titarchuk,
1995 [CT95]), for the first time, does not require any ad hoc Compton 
cloud or magnetic coronae to explain the power 
law components of black hole spectra. Here, one computes the properties 
of the so called `Compton cloud' self-consistently since it is a part of the 
inflow itself. 

In C96b, it is shown that the variation of the flow velocity
(and hence the flow density) close to the black hole is similar 
whether the shock forms or not (Fig. 7a of C96b). Hence, {\it qualitative} 
spectral properties of the TCAF Model does not seriously depend on whether
the shocks actually form in either or both of the Keplerian 
(which also becomes sub-Keplerian close to the hole) and the sub-Keplerian 
components. However, spectral properties {\it do  depend} upon 
whether the accretion flow is of single component (such as one Keplerian
flow becoming entirely sub-Keplerian close to the hole) or two components 
(where the original Keplerian disk plus companion winds are segregated into 
Keplerian and sub-Keplerian components very far away before being mixed 
into a single sub-Keplerian component near the hole). This will be 
demonstrated below. In this context, it is to be remembered that the 
sub-Keplerian flow can really be made up of two distinct components
(Fig. 1): one passes through the inner sonic point (forming the
giant disk; the solutions in g11-g41 grids of Fig. 2a of C96b)
and the other forms a shock (forming the centrifugally supported
dense region, the solution in grid g12-g42 of Fig. 2a of C96b). The central
Keplerian disk comes form the grid g13-g14 of the same Figure.

\section{Observational Properties of Multi-Component Advective Flows.}
%{\parfillskip=0pt 
Based upon above theoretical unstanding of the properties of advective flows,
CT95 pointed out that the accretion on most compact objects may be taking 
place in two components: one is of higher viscosity, predominantly Keplerian 
(Disk Component) and is extended till around $x_K \sim 10x_g$  or less
if the accretion rate is high enough to keep it thermally and viscously 
stable, otherwise $x_K$ could be higher (see also, ETC96). Keplerian 
region of the disk component supplies soft photons. The other component 
(Halo Component) is predominantly sub-Keplerian (which is originated from 
the Keplerian disk far away and is contributed by companion winds, if 
present, in the case of a galactic black hole and by winds from numerous 
stars in the case of a supermassive black hole.).  This component radiates 
inefficiently and therefore is hot ($\sim$ virial temperature; see  Rees, 
1984) and together with $x<x_K$ of the disk component they supply hot 
electrons which in turn energize intercepted soft photons (determined by the 
disk component) to produce the  hard component. The extent to which 
electrons cool is determined by the accretion rates in these two components. 
At least three important variations of this Model is recognized (Chakrabarti,
1997 [C97]): TCAFM1-- 
In this case, the halo component forms a strong shock behind the centrifugal 
barrier: $x_s\sim 10x_g$ and puffs up 
and mixes up with the disk component at $x<x_s$.
TCAFM2-- The halo component does not form a shock or 
forms only a weak shock but still feels the centrifugal barrier as in TCAFM1.
The results in these two models are similar. 
TCAFM3-- The halo component is completely devoid of angular momentum. 
The disk component deviates from a Keplerian disk at $x_K$. 
For $x<x_K$ these components mix as before. In this case, the absence of 
centrifugal barrier reduces the optical depth of the region $x<x_K$ and
it is easy to cool this region even at a low disk rate. 
A corollary of these Models is a single component model SCAFM, where
the sub-Keplerian component rate is so low that it is practically non-existent.
In SCAFM, soft photons of the Keplerian region may or may not cool the hot electrons
of its own sub-Keplerian region (for $x<x_K$) very effectively depending on $x_K$
and the spectra remains soft in most of the parameter space. Also, in this
case, the hard and soft components are always anticorrelated while observations
suggest that very often they behave independently. In CT95, TCAFM1 is 
extensively studied while other possibilities are also mentioned (see also ETC96). 
More detailed study of these models are in C97. 
Note that the generalized disk (Fig. 1) of the 1990s 
(and hopefully of the future) is really a natural combination of purely 
advecting Bondi flow of the 1950s
and purely rotating Keplerian disks of the 1970s. In Fig. 4 we show the basic
difference in the soft state spectra of neutron stars and black holes. In the soft state,
the disk rate is large and emitted soft photons completely cool the inner quasi-spherical
sub-Keplerian region. The inner boundary property on the horizon of a black hole causes the
cool (but rushing with velocity comparable to the velocity of light) electrons 
to Comptonize a fraction of these soft photons due to direct momentum transfer 
(as opposed to random 
\begin{figure}
\vbox{
\vskip -0.6cm
\centerline{
\psfig{figure=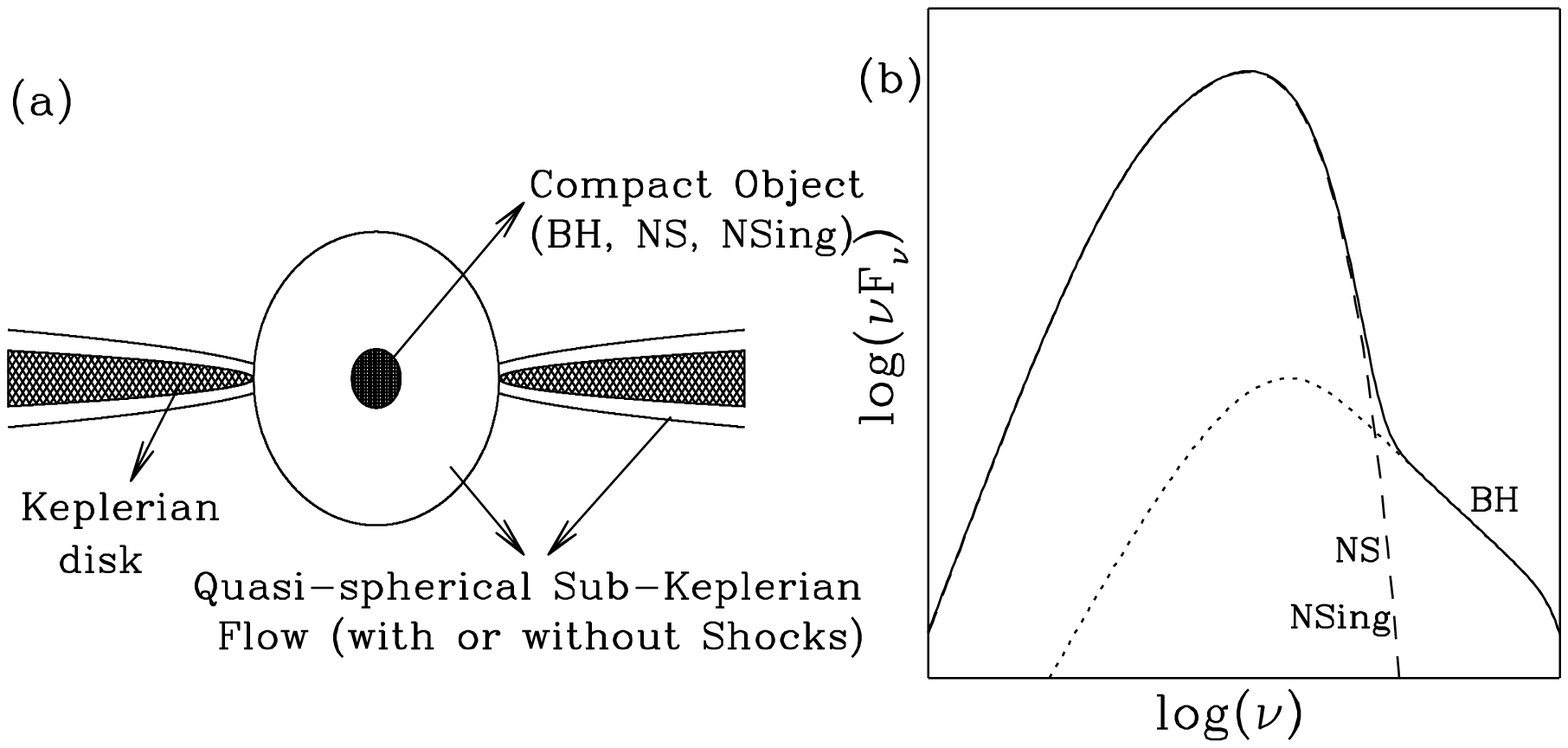,height=13.0truecm,width=13.0truecm,angle=0}}}
\vspace{-7.0cm}
\noindent {\small {\bf Fig. 4}: (a) The idealization of Advective flow
model of Fig. 1 and (b) its spectra in soft states. 
Soft states of neutron stars (NS), 
black holes (BH) and naked singularities (NSing) are distinguished by the 
presence or absence of the weak hard tail component due to bulk motion 
Comptonization (CT95).}
\end{figure}
%\par }
momentum transfer in thermal Comptonization) and produces
a weak hard tail component with photon index $ \sim 2.5-3$. While 
on a neutron star such a hard component must be missing since the flow 
has to slow down on the inner boundary. Just for completeness, we wish to
point out that a 
naked singularity (NSing, with inner boundary at $x \sim 0$) also should not 
have this weak hard tail since extremely dense advecting matter close to the
singularity would carry all such hard photons inwards. This feature is 
because the absorbing boundary is at $x=0$ 
rather than at $x=1.5$ (CT95; also see, Chakrabarti \& Sahu, 1997). 
Note that due to thermal Comptonization when the soft state is produced at 
around ${\dot m}_d \sim 0.1-0.3$
the energy spectral index can also be around $1.5-2.0$, but the power law is 
not extended till several hundred keV as in bulk motion Comptonization.
(Occasionally, this soft state without the extended power law
is simply called the `soft state' while that with an extended  power
law which is called `very soft state'; see Grebenev et al, 1994.)
Note also when relativistic corrections are added the
cutoff by the bulk motion Comptonization occurs at $4 m_e c^2 / {\dot m}$
(Titarchuk, 1997) rather than at $m_e c^2 /{\dot m}$ as in CT95. 
Thus the extended power law due to the bulk 
motion Comptonization could be as high as $\sim 1 MeV$. Indeed all the 
observed black hole candidates have this extended power law.
It is possible that the power law component may actually exist
beyond $\sim MeV$ is the soft state. This may be due to synchrotron radiation.
Magnetic field could be very strong in soft states due to higher
Keplerian disk rate (if equipartition is assumed). If this interpretation
is correct, then the power law component should not be extended 
in hard states till several MeV (since the magnetic field would be 
weaker due to equipartition with smaller disk rate). 

Another important prediction of our two component model is that 
during the state transition, the soft component may change
rather abruptly, while the hard component may change very slowly.
This is because the soft component produced by the Keplerian
disk which changes in a smaller time scale due to 
higher viscousity while the hard component will change in a smaller 
time scale due to the lower viscosity in the sub-Keplerian 
component. A factor of ten in these time scales is expected 
from this simple consideration alone. Indeed the recent
RXTE observation of Cyg X-1 shows precisely this behavior (Zhang et al.
1997 [Z97]). Furthermore,
since the transition from hard state to soft state in our model is due to the
sudden increase in the Keplerian rate, (i.e., viscosity 
--initiated by capturing of blobs of magnetic
fields for instance, which must take its advection time before
viscosity goes down again and the
hard state is resumed), the net luminosity may increase during
the soft state even if the total rate is fixed. 
This is because the efficiency of emission
from the sub-Keplerian component is not generally high 
(it depends on the number of soft photons supplied) in the hard state.
This has also been seen in the recent observation of  Cyg X-1 (Z97).

%{\parfillskip=0pt 
Fig. 5 shows examples of spectral transitions in 
black hole candidates in all the three models described above. We choose here
$M^*_{BH}=1M_\odot$, which after correction due to spectral hardening 
(Shimura \& Takahara, 1995), roughly corresponds to a mass of 
$M_{BH}\sim 3.6 M_\odot$. All the rates are in units of Eddington
rate. In Fig. 5(a), we consider three disk rates ${\dot m}_d 
=0.3,\ 0.05,\ 0.0005$ but the same halo rate ${\dot m}_h=1$. Solid, long-dashed
and short-dashed curves are for strong shock Model (TCAFM1), 
weak or no-shock Model (TCAFM2) and zero angular momentum halo  Model 
(TCAFM3). For a set of (${\dot m}_d, {\dot m}_h$), the
spectrum is hardest for TCAFM1 and softest for TCAFM3. This is expected since
the emission region has the highest optical depth when shocks are stronger. 
SCAFM always produces soft states for these parameters.
It can produce hard state in extreme parameter range. However,
in that case during the state transition, the hard and soft components
would always change in the same time scale which is not observed (Z97).
In Fig. 5(b), we show the comparison of energy spectral 
index $\alpha$ (where $F[E] \sim E^{-\alpha}$) for these models
as functions of ${\dot m}_d$. In all these models, spectra becomes soft
even when the disk rate is much below Eddington rate. In the case of 
supermassive black holes, the behavior is very similar
as the electron temperature of the sub-Keplerian region 
is very insensitive to the central mass ($T_e \propto M_{BH}^{0.04}$). 
At a high accretion rate, the bulk motion Comptonization produces 
weaker hard tail (CT95). Its behavior is independent 
of any model and depends mainly on the optical depth in the last few 
Schwarzschild radii outside the horizon. In the long dashed region of the 
convergent flow curve, both power laws due to thermal and bulk motion 
Comptonizations are expected in the observed
spectra. In Fig. 5(c), the dependence of the spectra on the location where the
flow deviates from a Keplerian disk ($x_K$) is shown using TCAFM3.
Fig. 5(d) shows the corresponding variation of the spectral index.
As described in CT95 and C96b, 
this variation of $x_K$ could be simply due to the viscosity variation 
in the flow (see, Fig. 3 above) and therefore such
\begin{figure}
\vbox{
\vskip -0.6cm
\centerline{
\psfig{figure=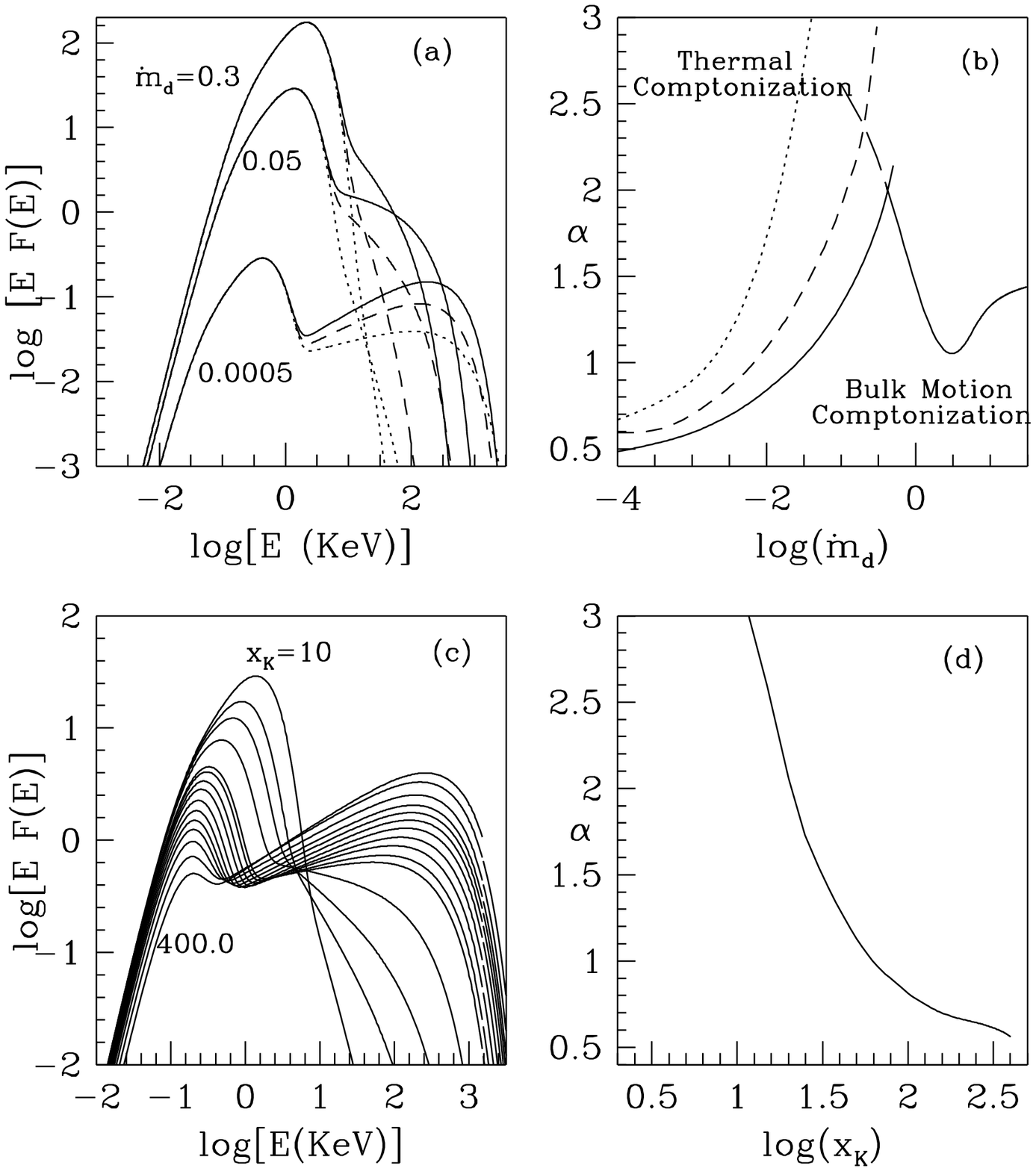,height=13.0truecm,width=13.0truecm,angle=0}}}
\vspace{-2.5cm}
\noindent {\small {\bf Fig. 5:} 
Model dependence of spectral properties. (a) Solid, long-dashed and short-dashed
curves are for TCAFM1, TCAFM2 and TCAFM3 respectively. (b) Spectral indices
of the corresponding models as functions of ${\dot m}_d$ are drawn.
(c) Result of TCAFM3 (${\dot m_d}=0.05,\ {\dot m_h}=1.0$). 
Spectrum changes from hard state to soft state as the Keplerian disk
approaches the black hole due to increase in $\alpha$ and ${\dot m}_d$
(as probably is the case in the rising phase of a novae outburst). 
(d) Changes in spectral index with $x_K$.}
\end{figure}
\noindent variation in the spectra is expected in viscous time scale,
specially during the rising phase of a novae outburst which is presumably 
induced
by an enhancement in viscosity (see, Cannizzo 1993 and references therein) at
the outer edge of the Keplerian disk. Figs. 5(c-d) were drawn for ${\dot m}_d=0.05$
and ${\dot m}_h=1.0$. Indeed, rising light curves in X-ray novae
derived from this consideration is remarkably similar to what is observed
(C97). Just as the initial outburst of an X-ray novae could be understood
by increase in the Keplerian rate (which is converted from the sub-Keplerian
matter as viscosity increases; see Chakrabarti \& Molteni, 1995) 
and consequent decrease of the inner edge of 
the Keplerian disk and/or the decay of the X-ray novae can also be understood
from the (primarily exponential) decay of the Keplerian rate. In reality both 
rates must change and the relative abundances of photons and electrons will 
determine whether a X-ray novae will even change its state from soft to hard
(as in GS1124-68 and GRS2000+25) or will remain hard for ever (as in 
GS2023+338, GRO J1655-40; see Grebenev et al, 1994). Decent spectral fits of 
GS2000+25 and GS1124-68 are presented in C97. Such transitions are best 
understood by numerical simulations of viscous advective flow together with 
radiative transfer which will be done in near future.

The quasi-periodic oscillations of black hole candidates are examples of
`puzzles' which are most naturally resolved within the framework of the
TCAF model. As we have already discussed above, there seems
to be two kinds of oscillations: in the case where the stable shock 
condition is not satisfied, but still a high entropy solution is
present (inner sonic point), the flow still forms an unstable oscillating
shock. The typical frequency is  a function of specific angular
momentum of the flow close to the inner edge (which in turn depends
on viscosity, see, C96b). Generally, the time period
is $4-6000 (2GM/c^3) \sim 4-6 (M/M_\odot) \times 10^{-2}$s (RCM97). 
In this case, the frequency of oscillation
may be somewhat insensitive to the accretion rate of the flow,
since the boundary of the region $NSA$ of Fig. 2 depends on specific
angular momentum and specific energy (which contains the information
of the temperature of the disk, which in turn, depends on 
cooling efficiency, and hence the accretion rate). Such QPOs should be
seen in sudden flare events when the specific energy temporarily
becomes positive (Fig. 2) and the flow parameters are
within $NSA$ or $NSW$. In a second kind of QPO, the infall time scale
in the post-shock flow (or, the flow crossing time scale in the centrifugally 
supported dense region) must be comparable to the cooling time scale (MSC96)
which also becomes the timescale of periodicity.
In this case, once the oscillation starts, the frequency must
increase roughly proportional to the accretion rate, if the two-body
processes are responsible (as in bremsstrahlung cooling). Otherwise,
in hard states during thermal Comptonization, the frequency
may remain vertually unchanged (since the Compton cooling time scale
remains roughly constant at low disk accretion rate). As the source
approaches the soft state, the Comptonization cooling behaves
highly non-linearly (see, the plot of $\alpha$ in Fig. 5b)
and hence the frequency should go up, initially linearly, but then
much faster as the Keplerian rate comes closer to the Eddington rate
(${\dot m}_d \sim 0.1-0.3$) pretty much like the variation of $\alpha$
as shown in Fig. 5b.

One point to note in this context is that it is possible to find 
a large number of physical mechanisms which may be able to explain
the {\it frequencies} of QPOs. However, the reason why we believe that 
the mechanism of oscillations as we mentioned above is operating
is that it is the only way known in the literature where the
oscillation amplitude could be as high as $10-50\%$ in normal circumstances,
and possibly even higher where the actual accretion on the compact object
is impeded (MSC96). The reason is that the hard X-ray
emiting region itself is dynamically `breathing' and intercepting the
variable number of soft photons, modulating the hard X-ray in the
process. The soft X-ray must be modulated by very less amount
(as is observed) since the variation in the inner edge of the
Keplerian component by a few ($1-3$) Schwarzschild radius
does not constitute drastic change in the luminosity from the Keplerian
disk. One way to check if this process is operating is to
monitor the peak frequency and the luminosity of 
the soft X-ray bump, both of which should
be anti-correlated with the luminosity of the hard X-ray. Of course,
it is not necessary that the shock forms in the sub-Keplerian
flow exactly where the Keplerian disk ends (Fig. 1) in which case,
such correlation is difficult to predict.
Other mechanisms in the literature can produce variations
of only a fraction of percentage.

\section{Concluding Remarks}

The understanding of accretion processes on black holes has undergone a major 
revision in recent years when the solutions of advective disks are
taken into considerations. It is now proven beyond doubt that the
sub-Keplerian flow plays a major role in shaping the stationary
and non-stationary spectra of black holes and neutron stars.
Spectral behavior of X-ray novae and other black hole candidates 
(see, e.g., ETC96 for LMC X-3, Crary et al. 1996; Z97
for Cyg X-1), universal presence of the weak hard tail 
in very soft states of black hole candidates, diverse observations
such as quiescent states to rising phases of black hole candidate novae, 
soft to hard transitions, pivoting property of the spectra, quasi-periodic 
oscillations (including observed large amplitude modulations) are naturally 
explained by TCAF models without invoking any additional unknown components.
Though we did not include magnetic fields explicitly, existence of small
fields as generated by say, Balbus-Hawley instability (Balbus \& Hawley,
1994), cannot affect our results. 

We already mentioned the importance of sub-Keplerian flows in the disks.
It is perhaps no accident that the sub-Keplerian flows were found to be
more effective in formation and collimation of cosmic radio jets
observed in active galaxies (Chakrabarti \& Bhaskaran, 1992). In active
galaxies, the transition of states (which occur in viscous time
scales) would take thousands of years! Thus, generally the
spectra is also found to remain in the same `state' for long time.
Nevertheless, there are several cases where the hard component has
been seen with a energy spectral slope of $1.5$ (e.g., Arnaud et al. 
1985). These indications, together with QPOs with periodicity of 
order of days (Halpern \& Marshall, 1996) may actually vindicate the 
validity of the advective disk model around both
galactic and extra-galactic black hole candidates.

\acknowledgments

The author acknowledges helpful discussions with L. Titarchuk and S. Nan Zhang.

\newpage

To appear in the Proceedings of the Golden Jubilee Meeting(Aug. 12th-16th, 1996)
of `Perspective of High Energy Astrophysics' of TIFR. Eds. P. Agarwal et al., 1997.

Author's Address to which Related reprints/Preprints could be sent is:

Prof. S.K. Chakrabarti, S.N. Bose National Center for Basic Sciences,
JD Block, Salt Lake, Calcutta - 700091, INDIA


\begin{references}
\reference Abramowicz, M.A., Czerny, B., Lasota, J.P. \& Szuzkiewicz, E. 1988,
ApJ, 332, 646
\reference Arnaud, K. et al. 1985, \mnras, 117, 105
\reference Balbus, S.A. \& Hawley, J.F. 1991, \apj 376, 214
\reference Bondi, H. 1952, \mnras, 112, 195
\reference Cannizzo, J. 1993 in Accretion Disks in Compact Stellar Systems J. C.
Wheeler, World Scientific: Singapore
\reference Chakrabarti, S.K. 1989, \apj, 347, 365 [C89]
\reference Chakrabarti, S.K. 1990a, \mnras, 243, 610 [C90a]
\reference Chakrabarti, S.K. 1990b, Theory of Transonic 
Astrophysical Flows, World Scientific: Singapore [C90b]
\reference Chakrabarti, S.K. 1996a, Phys. Rep, 266, 229 [C96a]
\reference Chakrabarti, S.K. 1996b, \apj, 464, 664 [C96b]
\reference Chakrabarti, S.K. 1996c, \mnras~ (Nov. 1st issue) [C96c]
\reference Chakrabarti, S.K. 1996d, \apj~ (Nov. 1st issue)  [C96d]
\reference Chakrabarti, S.K. 1997, \apj~ (in press) [C97]
\reference Chakrabarti, S.K. \& Bhaskaran, P. 1992, \mnras, 255, 255
\reference Chakrabarti, S.K., \&  Molteni, D. 1993, \apj, 417, 671 
\reference Chakrabarti, S.K. \& Molteni, D. 1995, \mnras, 272, 80
\reference Chakrabarti, S.K. \& Sahu, S. 1997, \astap ~  (in press)
\reference Chakrabarti, S.K. \& Titarchuk, L. G. 1995, \apj, 455, 623 [CT95]
\reference Crary, D.J. et al. \apj, 1996, 462, L71
\reference Dotani, Y. 1992 in Frontiers in X-ray Astronomy, Tokyo: Universal 
Academy Press, Y. Tanaka \& K. Koyama 152
\reference Ebisawa, K., Titarchuk, L. \& Chakrabarti, S. K., 1996, \pasj, 48, 1 [ETC96]
\reference Grebenev, S.A., Sunyaev, R.A. \& Pavlinsky, M.N., 1994, 
in Advances in Space Res., Proc. 30th COSPAR Scientific Assembly.
\reference Halpern, J. \& Marshall, H. L. 1996, \apj, 464, 760
\reference Molteni, D., Lanzafame, G., \& Chakrabarti, S. K. 1994, ApJ, 425, 161
\reference Molteni, D., Sponholz, H. \& Chakrabarti, S. K. 1996, \apj, 457, 805 [MSC96]
\reference Molteni, D., Ryu, D. \& Chakrabarti, S. K. 1996, \apj, (Oct 10th issue)
\reference Novikov, I. \&  Thorne, K. S. 1973, in Black Holes, 
C. DeWitt and B. DeWitt, Gordon and Breach: New York
\reference Papadakis, I. E. \& Lawrence, A. 1995, \mnras, 272, 161
\reference Rees, M.J. 1984, Ann. Rev. Astron. Ap., 22, 471
\reference Ryu, D., Chakrabarti, S.K. \& Molteni, D. 1997, \apj, (Jan. 1st) [RCM97]
\reference Shakura, N. I. \& Sunyaev, R. A. 1973, \astap, 24, 337 [SS73]
\reference Shimura, T. \& Takahara, F. 1995, \apj, 445, 780
\reference Titarchuk, L.G. 1997 Proc. 2nd INTEGRAL Workshop 
``The Transparent Universe" Eds. C. Winkler et al. (in press).
\reference Zhang, S.N. et al. 1997, \apj ~ (submitted)
\end{references}
\end{document}